\documentclass[conference]{IEEEtran}
\IEEEoverridecommandlockouts
\usepackage{cite}
\usepackage{amsmath,amssymb,amsfonts}
\usepackage{algorithmic}
\usepackage{graphicx}
\usepackage{textcomp}
\usepackage{xcolor}
\def\BibTeX{{\rm B\kern-.05em{\sc i\kern-.025em b}\kern-.08em
    T\kern-.1667em\lower.7ex\hbox{E}\kern-.125emX}}
\begin{document}

\newcommand\blfootnote[1]{%
  \begingroup
  \renewcommand\thefootnote{}\footnote{#1}%
  \addtocounter{footnote}{-1}%
  \endgroup
}

\title{Overview of Test Coverage Criteria for Test Case Generation from Finite State Machines Modelled as Directed Graphs\\
}



\author{\IEEEauthorblockN{Vaclav Rechtberger}
\IEEEauthorblockA{\textit{Dept. of Computer Science, FEE} \\
\textit{Czech Technical University in Prague}\\
Prague, Czechia \\
rechtva1@fel.cvut.cz}
\and
\IEEEauthorblockN{Miroslav Bures}
\IEEEauthorblockA{\textit{Dept. of Computer Science, FEE} \\
\textit{Czech Technical University in Prague}\\
Prague, Czechia \\
miroslav.bures@fel.cvut.cz }
\and
\IEEEauthorblockN{Bestoun S. Ahmed}
\IEEEauthorblockA{\textit{Dept. of Mathematics and Computer Science} \\
\textit{Karlstad University, Sweden \&} \\ Czech Technical University in Prague \\
bestoun@kau.se}
}


\maketitle

\begin{abstract}

Test Coverage criteria are an essential concept for test engineers when generating the test cases from a System Under Test model. They are routinely used in test case generation for user interfaces, middleware, and back-end system parts for software, electronics, or Internet of Things (IoT) systems. Test Coverage criteria define the number of actions or combinations by which a system is tested, informally determining a potential "strength" of a test set. As no previous study summarized all commonly used test coverage criteria for Finite State Machines and comprehensively discussed them regarding their subsumption, equivalence, or non-comparability, this paper provides this overview. In this study, 14 most common test coverage criteria and seven of their synonyms for Finite State Machines defined via a directed graph are summarized and compared. The results give researchers and industry testing engineers a helpful overview when setting a software-based or IoT system test strategy.
\end{abstract}



\section{Introduction}

\color{blue}\textbf{Paper accepted at the 5th INTUITESTBEDS (International Workshop on User Interface Test Automation and Testing Techniques for Event Based Software), part of the 15th IEEE International Conference on Software Testing, Verification and Validation (ICST) 2022, April 4 - 13, 2022, https://icst2022.vrain.upv.es/ }
\color{black}

\hspace{3mm}

Finite State Machines (FSMs) are employed as a favorite notation in system modeling and testing, as they describe a natural aspect of a wide variety of systems. Systems themselves, their modules or parts, user interfaces, or data objects processed by a system naturally switch from state to state. Hence, FSMs are one of the fundamental modeling options for Model-based Testing generation of test cases. The FSM-based test case generation is applicable on a broad spectrum of systems to test, including software \cite{ammann2016introduction}, electronics \cite{yin2012real}, networking systems \cite{bosik1991finite}, and its combinations as recent popular Internet of Things (IoT) \cite{tanabe2020model}. Regarding the system level, FSM-based test case generation can be utilized in user interface testing, middleware level, or lowers levels of a System Under Test (SUT) universally. \blfootnote{Bestoun S. Ahmed is also with Dept. of Computer Science, FEE, Czech Technical University in Prague.}



The SUT model can be based on a directed graph or a grammar notation in FSM testing. The core of definitions based on a directed graph is a directed graph $\mathcal{G} = (V,E,v_s,V_e)$ such that $V \neq \emptyset$ is a finite set of vertices representing FSM states and $E \subseteq N \times N$ is a nonempty set of edges $e \in E$ representing FSM transitions. Further, $v_s \in V$ is start state of the state machine, $V_e \subset V$ is a set of end states of the state machine \cite{wang2019formal,hopcroft2001introduction,ammann2016introduction}. Alternative SUT models based on various extensions of directed graphs or other structures as grammars are available. However, in this paper, we used the $\mathcal{G}$ model as it is a common widely used model in the literature. 

A test path $t$ is a path in $\mathcal{G}$. The test path does not have to start in $v_s$ and end in an end vertex from $V_e$. Finally, $T$ denotes a set of test paths.


Coverage criterion generally defines a property or a set of properties $T$ has to satisfy. Coverage criteria are used to express the "strength" of the test paths, which usually implies the number of steps or various combinations to be performed. However, at this point, it is worth mentioning that this notion of "strength" is rather intuitive and widely used by testing practitioners. Still, its use is sometimes discouraged by the research community in system testing. The number of steps or combinations exercised is an exact property of $T$ that can be discussed formally, but the potential of $T$ to detect the defects is a more complex issue. At least, thoroughness of the particular execution of the individual test steps plays a role here, and the tester's initiative and analytical thinking during the test execution. This applies especially when test paths are defined more high-level (as $T$ in this study), and testers have some liberty to try more combinations or situations in the individual test steps.

More test steps or combinations contribute to more "strong" tests, but it's not the only factor that influences the final potential of tests to find some defects. Despite this limitation, the coverage criteria are a powerful tool to define the test paths' properties and potential "strength." Ammann and Offutt have done some overview work for test coverage criteria related to path-based testing \cite{ammann2016introduction}. Some of these criteria are applicable to FSM testing as well. However, we have not found a comprehensive overview of test coverage criteria dedicated directly to FSM testing. Instead, only individual works discussing some of the criteria and their relations can be found \cite{Watson1996StructuredTA,5676826,1245299,792624,Mukherjee2016,Lun2019}.

To this end, we provide an overview of the most common criteria for $\mathcal{G}$ in this paper. Besides this overview, we compare these criteria in terms of their mutual relations. At this point, we consider subsumption, equivalence, and the fact that two coverage criteria are incomparable (relations are further defined in Section \ref{sec:subsumption_of_coverage_criteria}). 

This paper focuses on FSM solely, and Extended Finite State Machines (EFSM) that are also used in system testing are not considered directly. Most of the test coverage criteria discussed in this paper may be applicable; however, this limit has to be taken into account. Also, fault coverage producing negative test cases for illegal or missing states and transitions (e.g., \cite{mariano2019comparing}) are not discussed in the paper. 



\section{The common test coverage criteria for graph-based definition of FSM}
\label{sec:coverage_criteria_for_graph}

In this section, we provide an overview of the 14 most common test coverage criteria related to FSM modeled as $\mathcal{G}$, based on the literature search and our industrial praxis.





\subsection{Node Coverage (NC)} 
$T$ satisfies NC, when each $v \in V \in \mathcal{G}$ is present in at least one $t \in T$. In the literature NC is also alternatively denoted as \textit{All States Coverage} \cite{ammann2016introduction,ModelBasedTestCasesGenerationForOnboardSystem}. Considering the number of $T$ steps, this criterion is the weakest of all criteria discussed in this section  \cite{CoverageCriteriaForStateTransitionTestingAndModelCheckerBasedTestCaseGeneration}.

\subsection{Edge Coverage (EC)} 
$T$ satisfies EC, when all edges (referred also as the transitions) $e \in E \in \mathcal{G}$ are present in at least one $t \in T$. This criterion is usually referred also as \textit{0-Switch coverage} or \textit{All Transitions Coverage} \cite{ammann2016introduction,ModelBasedTestCasesGenerationForOnboardSystem,CoverageCriteriaForStateTransitionTestingAndModelCheckerBasedTestCaseGeneration,Heimdahl2004TestSuiteReductionForModelBasedTests}.


The EC can also be discussed in the program source code execution context. When the source code structure is modeled as FSM in the way that code statements are modeled as only $\mathcal{G}$ edges, \textit{Statement Coverage} criterion usually discussed in this context would be equivalent to EC. 

\subsection{Branch Coverage (BC)} 
To satisfy BC, each branch of the code must be present in at least one $t \in T$ \cite{AnApproachToProgramTesting}. When employing $\mathcal{G}$ to model the code structure, the branch is defined as a path that satisfies the following conditions:

\begin{enumerate}
    \item The first edge of the branch starts from $v_s \in \mathcal{G}$ or from a $v \in \mathcal{G}$ that represents a decision in the code structure,
    \item the last edge of the branch ends in some $v \in V_e \in \mathcal{G}$ or in a $v \in \mathcal{G}$ that represents a decision in the code structure, and, 
    \item there is no other decision point in this path.
\end{enumerate}





\subsection{Edge-Pair Coverage (EPC)} 
$T$ satisfies EPC when each path that consists of two adjacent edges $e \in E$ must occur at least once in at least one $t \in T$ \cite{ammann2016introduction,Li2009AnExperimentalComparisonOfFourUnitTestCriteria}. EPC is also mentioned in the literature as \textit{All Transition Pairs Coverage} and \textit{1-Switch Coverage} \cite{CoverageCriteriaForStateTransitionTestingAndModelCheckerBasedTestCaseGeneration}. Moreover, modifications of this criterion can be found in the literature:

\begin{itemize}
  \item EPC with an additional constraint that the transitions must not be adjacent (differently to standard EPC criterion)
  \cite{CoverageCriteriaForStateTransitionTestingAndModelCheckerBasedTestCaseGeneration}.
  \item EPC with an additional constraint that any pair must not contain two same edges \cite{ANewApproachToGeneratingHighQualityTestCases}.
\end{itemize}

\subsection{All Path Coverage (APC)} 

$T$ satisfies APC when each possible path in $\mathcal{G}$ is present in $T$. When $\mathcal{G}$ contains cycles, it is impossible to satisfy this criterion - if there is at least one cycle, the number of possible paths is infinite \cite{ammann2016introduction}. APC is also mentioned in the literature as \textit{Complete Path Coverage} \cite{ammann2016introduction}.

\subsection{Prime Path Coverage (PPC)} 
\label{sec:prime_path_coverage}
$T$ satisfies PPC, when each prime path of $\mathcal{G}$ is a sub path of at least one $t \in T$ \cite{ammann2016introduction,Li2009AnExperimentalComparisonOfFourUnitTestCriteria,Fazli2019ATimeAndSpaceEfficientCompositionalMethodForPrimeAndTestPathsGeneration}. A path is simple \textit{Simple path} when it does not contain any loop, except this path can be a cycle itself \cite{ammann2016introduction}. The \textit{Prime path} is a simple path that is not a subpath of any other simple path \cite{ammann2016introduction}.


\subsection{Specified Path Coverage (SPC)} 
$S$ is a set of specified paths that must be present in $T$. Then, $T$ satisfies SPC, when each $s \in S$ is present in at least one $t \in T$ \cite{ammann2016introduction}. The principle of this criterion is practically equivalent to the Test Requirements concept from the general path-based testing \cite{ammann2016introduction,bures2019employment,bures2019prioritized}.

\subsection{Round Trip Coverage (RTC)} 
RTC is a special case of PPC criterion covering the possible round trips in $\mathcal{G}$.  \cite{ammann2016introduction}. A \textit{Round trip path} is a prime path of nonzero length that starts and ends at the same vertex. \cite{ammann2016introduction}. There are two commonly used types of this criterion: \textit{Simple Round Trip Coverage} and \textit{Complete Round Trip Coverage}.

\subsection{Simple Round Trip Coverage (SRTC)} 
\label{sec:simple_round_trip_coverage}

SRTC is satisfied, when $T$ contains at least one round trip path for each reachable vertex $v \in V \in \mathcal{G}$ that starts and ends an round trip path in $\mathcal{G}$ \cite{ammann2016introduction}.


\subsection{Complete Round Trip Coverage (CRTC)} 
\label{sec:complete_round_trip_coverage}
CRTC is satisfied, when $T$ contains all possible round trip paths for each reachable vertex $v \in V \in \mathcal{G}$ \cite{ammann2016introduction,StateBasedTestsSuitesAutomaticGenerationTool,khalil2017OnFSMBasedTesting,khalil2017FiniteStateMachineTestingCompleteRoundTripVersusTransitionTrees}.

\subsection{Basis Path Coverage (BPC)}

$T$ satisfies BPC when $T$ is a maximal set of basis paths - linearly independent paths of $\mathcal{G}$ (paths that are not a linear combination of other paths in $\mathcal{G}$). The number of basis paths is equal to cyclomatic complexity \cite{McCabe1976AComplexityMeasure,Watson1996StructuredTA,ZAPATA2013256,Watson1996StructuredTA2,Wang2019BasisPathCoverageCriteriaForSmartContractApplicationTesting}. When applied at the source code level, BPC guarantees that every statement of the code is executed at least once in the tests \cite{McCabe1976AComplexityMeasure}.

\subsection{W Method Coverage (WMC)}
\label{sec:w_method_coverage}
The W Method Coverage Criterion proposed by Chow \cite{TestingSoftwareDesignModeledByFiniteStateMachines}.  It is based on the program \textit{Testing tree}, also referred to as the \textit{Transition tree}, on all paths of the transition tree that starts in the initial node and ends in a leaf node, denoted as \textit{P}, and characterization set denoted as \textit{W}. The breadth-first search algorithm generates the testing tree. The Characterization set is a set of input sequences, such that for each pair of states of the whole FSM, there is at least one sequence. By applying it to these states, we obtain two different output sequences. The Final Test Set is obtained as concatenation of each $p \in P$ by each possible input sequence $w \in W$. 

WMC is alternatively referred as \textit{Transition Tree Coverage} \cite{StateBasedTestsSuitesAutomaticGenerationTool}. In a few case studies, Chow showed that WMC criterion is stronger in its fault detection power than \textit{Branch Coverage}, \textit{1-Switch Coverage}, \textit{Boundary-Interior Coverage} and \textit{H-Language Coverage}. Here, the fault detection power is dependent on particular cases (presence of defects in the tested systems). It is practically impossible to define such a criterion precisely. We can define the only subsumption of coverage criteria, as discussed further in \ref{sec:subsumption_of_coverage_criteria}.


\subsection{N-Switch Coverage (NSC)}
In the industry, NSC is one of the most known and universal coverage criteria for FSMs, mainly for its presence in the TMAP Next methodology \cite{vroon2013tmap,pol2002software,aalst2008tmap}. $N$-Switch Coverage is satisfied when every combination of $N+1$ adjacent transitions (edges of $\mathcal{G}$) occur at least once in some $t \in T$ \cite{ConcurrentNSwitchCoverageCriterionForGeneratingTestCasesFromPlaceTransitionNets}. Two special cases of NSC are commonly used:

\begin{description}
  \item[0-Switch Coverage] \hfill \\ Equivalent to EC already defined above \cite{ConcurrentNSwitchCoverageCriterionForGeneratingTestCasesFromPlaceTransitionNets}.
  \item[1-Switch Coverage] \hfill \\ Equivalent to EPC already defined above \cite{ConcurrentNSwitchCoverageCriterionForGeneratingTestCasesFromPlaceTransitionNets}.
  

  
\end{description}

\subsection{Boundary-Interior Coverage (BIC)}
For all possible paths through a program (paths in $\mathcal{G}$), a finite set of classes $\mathcal{C}$ is determined when having two paths $p_1$ and $p_2$ that traverse $\mathcal{G}$ in the same way except in traversal of loops. They are placed in separate classes if:
\begin{enumerate}
    \item one of the paths enters the loop but does not repeat it, and the second path repeats this loop at least once,
    \item in both $p_1$ and $p_2$ we enter or leave the loop by a different entry or exit point,
    \item in both $p_1$ and $p_2$ we only enter the loop and traverse the loop by a different path, or,
    \item in both $p_1$ and $p_2$ we repeat the loop, and then we traverse the loop by a different path.
\end{enumerate}

$T$ satisfies BIC, when it contains at least one path from each $c \in \mathcal{C}$ \cite{Howden1975MethodologyForTheGenerationOfProgramTestData, GenerationOfIntegrationTestsForSelf-TestingComponents}.




\section{Subsumption of coverage criteria for graph-based Model}
\label{sec:subsumption_of_coverage_criteria}

In this section, we discuss a hierarchy of common test coverage criteria for a graph-based model of SUT as presented in Section \ref{sec:coverage_criteria_for_graph}. We use the term \textbf{subsume} with the following meaning: A test coverage criterion $C_1$ subsumes $C_2$ if each test set that satisfies $C_1$ will satisfy $C_2$ as well, and this relation is transitive.

From the test practitioner's viewpoint, we can informally say that $C_1$ is "stronger" than $C_2$. Also, $C_1$ produces more test combinations than $C_2$. But we cannot say that all test paths from a test set satisfying $C_2$ are present in a test set satisfying $C_1$.

\begin{figure*}
    \centering
    \includegraphics[width=18cm]{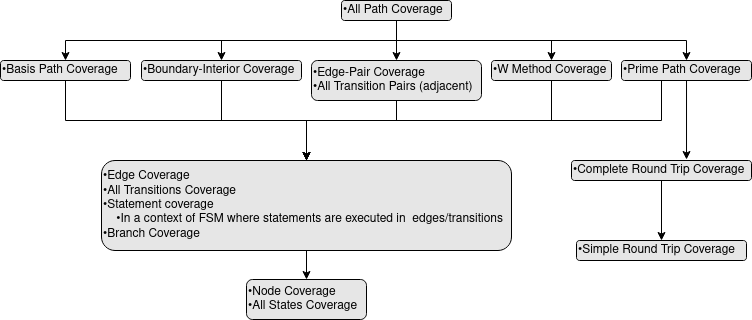}
    \caption{Hierarchy of common test coverage criteria for graph-based SUT model.}
    \label{fig:criteria_hierarchy}
\end{figure*}

Figure \ref{fig:criteria_hierarchy} presents the hierarchy of criteria presented in Section \ref{sec:coverage_criteria_for_graph} in terms of the subsume relation defined above. Suppose a test coverage criterion in rectangle A subsumes some other coverage criteria in rectangle B. In that case, this is depicted by an arrow that goes from rectangle A to rectangle B. Test coverage criteria in the same rectangle are equivalent.

In the following section, we explain the subsume relations captured in Figure \ref{fig:criteria_hierarchy}.


%


Besides \textit{subsume} relation defined above, we use the following terms. In the case that $C_1$ subsume $C_2$ and vice versa, we will say that $C_1$ is \textbf{equivalent} to $C_2$; this relation is transitive and symmetric. Furthermore, we say that $C_1$ is \textbf{incomparable} to $C_2$ if $C_1$ does not subsume $C_2$ and $C_2$ does not subsume $C_1$. Note that if we haven't compared some test coverage criteria in this section, it does not automatically mean that these two criteria are incomparable.

\subsection{Explanations of presented subsume relations}




\textbf{All Path Coverage criterion subsumes all other criteria.}

\textit{Explanation:} Since to satisfy \textit{All Path Coverage}, we need all paths of the graph to be traversed, this criterion automatically subsumes the other criteria in the hierarchy. All other criteria aim to reduce the number of test paths.

\textbf{There are not any criteria, that subsume Node Coverage.}

\begin{figure}
    \centering
    \includegraphics[width=1.2cm]{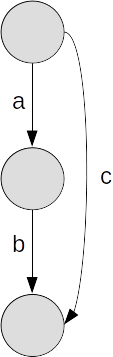}
    \caption{Graph for discussion of Node Coverage position in the hierarchy.}
    \label{fig:node_coverage_proof}
\end{figure}

\textit{Explanation:} \textit{Node Coverage} is at the very bottom of the hierarchy because this criterion does not even guarantee to tour all edges in a SUT model. This can be easily proven by a test set \textit{\{a-b\}} that satisfy \textit{Node Coverage} criterion in a graph in Figure \ref{fig:node_coverage_proof} but obviously \textit{Edge Coverage} criterion is not satisfied by this test set.

\textbf{Prime Path Coverage subsumes Complete Round Trip Coverage as well as Simple Round Trip Coverage.}

\textit{Explanation:}  Test sets that satisfy \textit{Complete Round Trip Coverage}, as well as \textit{Simple Round Trip Coverage}, contain only prime paths that are cycles, and a test set satisfying \textit{Prime Path Coverage} contains all prime paths in a graph.

\textbf{Complete Round Trip Path Coverage subsumes Simple Round Trip Path Coverage.}

\textit{Explanation:}  This relation has been proven by Ammann and Offutt \cite{ammann2016introduction}.

\textbf{Both Complete Round Trip Path Coverage and Simple Round Trip Path Coverage do not subsume any other coverage criteria discussed in Section \ref{sec:coverage_criteria_for_graph}.}

\textit{Explanation:} In the case of a graph with no cycle, both criteria do not require any path to traverse, hence, for a graph with no cycle, $P=\emptyset$ satisfies both \textit{Complete Round Trip Path Coverage} and \textit{Simple Round Trip Path Coverage}. But $P=\emptyset$ does not satisfy any other criteria discussed in Section \ref{sec:coverage_criteria_for_graph}.


\textbf{Branch Coverage subsumes Edge Coverage.}

\textit{Explanation:} The fact that "\textit{Branch Coverage} subsumes \textit{Edge Coverage}" will be obvious when we negate it as "There is a test set that satisfies \textit{Branch Coverage} and do not satisfy \textit{Edge Coverage}". This negated claim is not true because, in FSM, each edge is a part of exactly one branch.


\textbf{Edge Coverage subsumes Branch Coverage.}

\textit{Explanation:} The fact that "\textit{Edge Coverage} subsumes \textit{Branch Coverage}" will be obvious when we negate it as "There is a test set that satisfies \textit{Edge Coverage} but does not satisfy \textit{Branch Coverage}". When this claim is true, all edges of FSM would be traversed, but not every branch of this FSM. That would imply that each edge of each branch would be traversed, but still, not every branch of FSM is needed to be traversed, which is an obvious contradiction.


\textbf{Branch Coverage is equal to Edge Coverage.}

\textit{Explanation:} As explained above, \textit{Branch Coverage} subsumes \textit{Edge Coverage} and, concurrently, \textit{Edge Coverage} subsumes \textit{Branch Coverage}, hence we consider these two criteria as equal.


\textbf{Neither Node Coverage nor Edge Coverage subsume any of the Round Trip Coverages (Complete Round Trip Path Coverage and Simple Round Trip Path Coverage).}

\begin{figure}
    \centering
    \includegraphics[width=2.5cm]{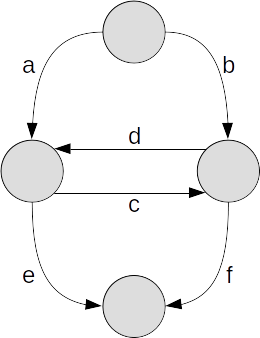}
    \caption{Example for discussion of relation between Node Coverage, Edge Coverage and Round Trip Coverages.}
    \label{fig:graph_rtc_not_comparable}
\end{figure}

\textit{Explanation:} A test set satisfying neither \textit{Node Coverage} nor \textit{Edge Coverage} would not satisfy both \textit{Round Trip Coverages} since \textit{Round Trip Coverages} requires to traverse the graph representing FSM in concrete way specified in Sections \ref{sec:simple_round_trip_coverage} and \ref{sec:complete_round_trip_coverage} but \textit{Node Coverage} and \textit{Edge Coverage} do not require to traverse the graph this way. For instance test set $\{a- c- f, b- d- f\}$ satisfy \textit{Edge Coverage} for graph in Figure \ref{fig:graph_rtc_not_comparable} but not any of the \textit{Round Trip Coverages}. 

\textbf{Edge Coverage and Node Coverage are incomparable with both Round Trip Coverage}

\textit{Explanation:} This is a consequence of the facts "Both Complete Round Trip Path Coverage and Simple Round Trip Path Coverage do not subsume any other coverage criteria discussed in Section \ref{sec:coverage_criteria_for_graph}" and "Neither Node Coverage nor Edge Coverage subsume any of the Round Trip Coverages (Complete Round Trip Path Coverage and Simple Round Trip Path Coverage)".


\textbf{Edge Coverage subsumes Node Coverage.}

\textit{Explanation:} Because each edge is defined by two nodes, all nodes must be traversed (to satisfy \textit{Node Coverage}) when all edges are traversed (to satisfy \textit{Edge Coverage}).


\textbf{Edge-Pair Coverage subsumes Edge Coverage.}

\textit{Explanation:} A test set satisfying \textit{Edge-Pair Coverage} criterion also satisfies \textit{Edge Coverage} because when the test paths tour all edge pairs, all edges must be toured by these paths as well.



\textbf{Edge Coverage does not subsume Edge-Pair Coverage}

\begin{figure}
    \centering
    \includegraphics[width=1.8cm]{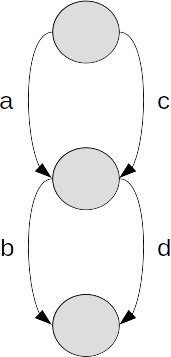}
    \caption{An example for discussion of relation between Edge Coverage and Edge-Pair Coverage.}
    \label{fig:edge-pair_coverage_proof}
\end{figure}

\textit{Explanation:} This fact can be demonstrated using an example. A test set \textit{\{a-b,c-d\}} satisfies \textit{Edge Coverage} for a graph presented in Figure \ref{fig:edge-pair_coverage_proof} but it does not satisfy \textit{Edge-Pair Coverage} since combinations \textit{a-d} and \textit{c-b} that would be required to be present in the test set to satisfy \textit{Edge-Pair Coverage} are missing there.

\textbf{Prime Path Coverage does not subsume Edge-Pair Coverage}

\begin{figure}
    \centering
    \includegraphics[width=6cm]{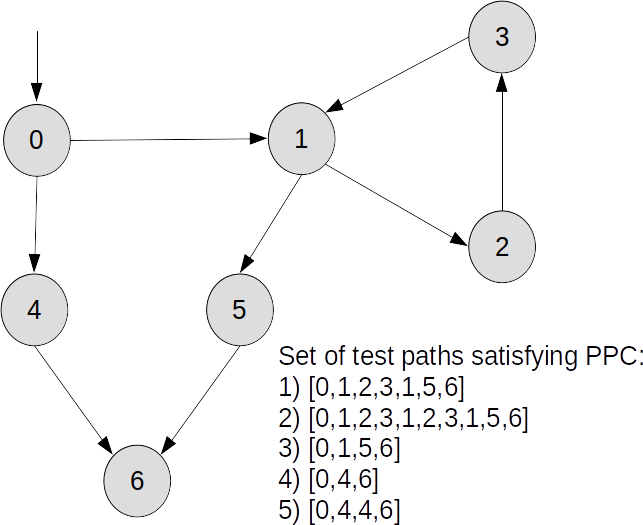}
    \caption{Prime Path Coverage example \cite{ammann2016introduction}.}
    \label{fig:example_from_ammann}
\end{figure}

\textit{Explanation:} This claim can be explained by the example for \textit{Prime Path Coverage} given in \cite{ammann2016introduction}. This example follows in Figure \ref{fig:example_from_ammann}.

In test paths that would satisfy \textit{Prime Path Coverage}, there is no sequence \textit{[4,4,4]}, which is needed to satisfy \textit{Edge-Pair Coverage}.

\textbf{Prime Path Coverage does not subsume Edge-Pair Coverage}

\begin{figure}
    \centering
    \includegraphics[width=1.8cm]{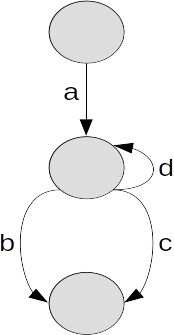}
    \caption{Example for relation between Prime Path Coverage and Edge-Pair Coverage.}
    \label{fig:ppc_imply_epc_not_truth_proof_example}
\end{figure}

\textit{Explanation:} For a graph in Figure \ref{fig:ppc_imply_epc_not_truth_proof_example}, prime paths are \textit{\{a-b,a-c,d\}}. In this set of prime paths, there are not all possible paths of length 2 (for instance \textit{d-b}) that would be required to be there to satisfy \textit{Edge-pair Coverage}. Hence, \textit{Prime Path Coverage} does not guarantee to cover all edge-pairs.

\textbf{Edge-Pair Coverage does not subsume Prime Path Coverage.}

\begin{figure}
    \centering
    \includegraphics[width=1.8cm]{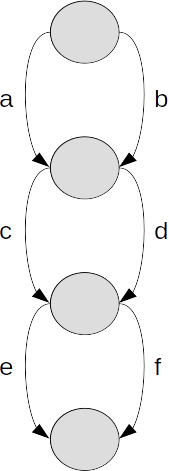}
    \caption{Example graph for discussion of relation between Edge-Pair Coverage and Prime Path Coverage.}
    \label{fig:graph_for_epc_stronger_ppc_disproof}
\end{figure}

\textit{Explanation:} This claim can be demonstrated using the following example. Figure \ref{fig:graph_for_epc_stronger_ppc_disproof} presents a graph for which a set of prime paths consists of all possible paths in this graph, in particular $\{x-y-z~|~x \in \{a,b\}, y \in \{c,d\}, z \in \{e,f\} \}$. Removing a path $a-c- e$, the resulting set will not satisfy \textit{Prime Path Coverage} since the removed path is prime, but this set still satisfies \textit{Edge-Pair Coverage}. When removing a test path $a-c-e$ that consists of two edge pairs, $a-c$ and $c-e$, we have to ensure, that these pairs are still covered by a reduced set to satisfy \textit{Edge-Pair Coverage}. And the pairs $a-c$ and $c-e$ are really covered by the reduced test set, the first pair is covered by path $a- c - f$ and the second pair is covered by $b- c - e$. 
Hence, \textit{Edge-Pair Coverage} does not subsume \textit{Prime Path Coverage}.

\textbf{Prime Path Coverage and Edge-Pair Coverage are incomparable.}

\textit{Explanation:} Because \textit{Prime Path Coverage} does not subsume \textit{Edge-Pair Coverage} and \textit{Edge-Pair Coverage} does not subsume \textit{Prime Path Coverage}, we consider these two criteria as incomparable.


\textbf{Basis Path Coverage subsumes Edge Coverage.}

\textit{Explanation:} It has been proven by Watson et al. that \textit{Basis Path Coverage} satisfies \textit{Edge Coverage}, but not vice versa \cite{Watson1996StructuredTA}.


\textbf{Basis Path Coverage does not subsume Edge-pair Coverage.}

\begin{figure}
    \centering
    \includegraphics[width=7cm]{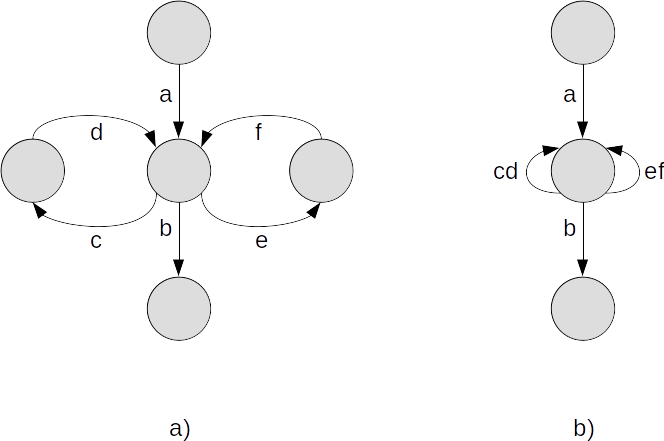}
    \caption{Examples for test coverage criteria discussion.}
    \label{fig:further_research_question_example_graphs}
\end{figure}

\textit{Explanation:} To document this relation, we can use graph \textit{a)} from Figure \ref{fig:further_research_question_example_graphs}. For this graph, a test set \textit{\{a-b, a-c-d-b, a-e-f-b\}} satisfies \textit{Basis Path Coverage} criterion, but in this test set, the edge pairs \textit{d-e} and \textit{f-c}, which would be needed to satisfy the \textit{Edge-pair Coverage} criterion, are missing.

\textbf{Edge-pair Coverage does not subsume Basis Path Coverage.}

\textit{Explanation:} This relation can be documented by the example graph \textit{a)} given in Figure \ref{fig:further_research_question_example_graphs}. A test set \textit{\{a-b, a-c-d-e-f-b, a-e-f-c-d-b\}} satisfies \textit{Edge-pair Coverage} but not the \textit{Basis Path Coverage} since you cannot obtain a required test path \textit{a-c-d-b} by a linear combination of the paths in the discussed test set \textit{\{a-b, a-c-d-e-f-b, a-e-f-c-d-b\}}.


\textbf{Basis Path Coverage is not comparable with Edge-Pair Coverage.}

\textit{Explanation:} This fact is a consequence of already discussed facts "\textit{Edge-pair Coverage} does not subsume \textit{Basis Path Coverage}" and "\textit{Basis Path Coverage} does not subsume \textit{Edge-pair Coverage}".



\textbf{Basis Path Coverage does not subsume Prime Path Coverage.}

\begin{figure}
    \centering
    \includegraphics[width=2.2cm]{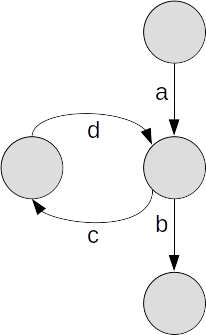}
    \caption{An example for discussion of the relationship between Basis Path Coverage and Prime Path Coverage.}
    \label{fig:graph_bpc_is_not_stronger_than_ppc}
\end{figure}

\textit{Explanation:} We document this fact using an example graph in Figure \ref{fig:graph_bpc_is_not_stronger_than_ppc}. A test set $\{a - b, a - c - d - b\}$ satisfies \textit{Basis Path Coverage} criterion, but this test set does not contain a prime path $d - c$ required to satisfy the \textit{Prime Path Coverage} criterion.

The next criterion to discuss is \textit{\textbf{W Method Coverage}}. As explained in Section \ref{sec:w_method_coverage}, in this method, test sets are generated using the \textit{Testing Tree} and set of inputs $W$. $W$ cannot be obtained only from $\mathcal{G}$ because it depends on FSM inputs and outputs. Hence, we can only say that there will be at least one input sequence for each state (except the final state) that triggers at least one adjacent transition from this state to another. Still, it is uncertain which one will be traversed in the case of states with more than one outgoing transition. A consequence of this fact is that each path (that starts in $v_s \in \mathcal{G}$ node and ends in a leaf node) obtained from the \textit{Testing Tree} will be appended with at least one edge (except the paths that end in a leaf node), but not necessarily with all possible edges.


\textbf{W Method Coverage does not subsume Edge-pair Coverage.}

\begin{figure}
    \centering
    \includegraphics[width=7cm]{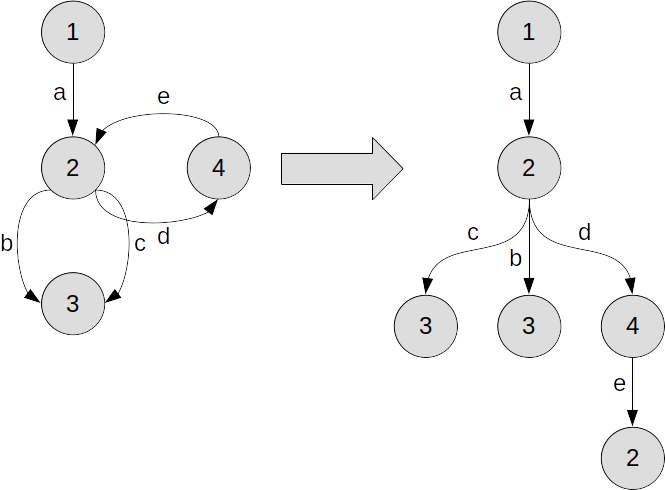}
    \caption{Example to discuss relation Graph and corresponding Testing tree.}
    \label{fig:wmc_graph_to_testing_tree}
\end{figure}

\textit{Explanation:} Let's use an example of a \textit{Testing Tree} used to generate a test set satisfying \textit{W Method Coverage} that is given in the right part of Figure \ref{fig:wmc_graph_to_testing_tree}. When we analyze paths in this \textit{Testing Tree} which start in an initial node and end in a leaf node, we cannot guarantee that all sub-paths of length 2, namely $e-b$, $e-c$ or $e-d$, that would be required to satisfy \textit{Edge-pair Coverage} will be traversed.

\textbf{W Method Coverage does not subsume Prime Path Coverage.}

\textit{Explanation:} In the \textit{Testing Tree} in the example given in Figure \ref{fig:wmc_graph_to_testing_tree}, a path $a - d - e$ will be appended at least by one of the adjacent edges $\{b, c, d\}$ but not necessary by all, so each of the prime paths $e - b$, $e - c$ or $e - d$ cannot be guaranteed to be covered to satisfy \textit{Prime Path Coverage} criterion.


\textbf{W Method Coverage does not subsume Basis Path Coverage.}

\textit{Explanation:} We use an example to document this fact. The basis paths set for graph in Figure \ref{fig:wmc_graph_to_testing_tree} is $\{a - b, a - c, a - d - e - b, a - d - e - c \}$ and with this set, \textit{Basis Path Coverage} is satisfied. However, \textit{W Method Coverage} cannot guarantee to cover both paths $e - b$ and $e - c$, from this set. Hence, \textit{W Method Coverage} cannot be satisfied by this test set.



\textbf{N-Switch Coverage with $N + 1 \geq (length~of~the~longest~path~in~\mathcal{G}$ is equivalent to All Path Coverage.}

\textit{Explanation:} Theoretical possible length of test paths satisfying \textit{N-Switch Coverage} where $N +1 \geq length~of~the~longest~path~in~\mathcal{G}$ is simply the same as a length of the longest path in $\mathcal{G}$. So, all paths in $\mathcal{G}$ will be present in a test set satisfying \textit{N-Switch Coverage} with $N + 1 \geq length~of~the~longest~path~in~\mathcal{G}$.



\textbf{N-Switch Coverage with $N + 1 \ge length~of~the~longest~prime~path~in~\mathcal{G}$ subsumes Prime Path Coverage.}

\textit{Explanation:} \textit{N-Switch Coverage} criterion with $N + 1 \ge length~of~the~longest prime~path~in~\mathcal{G}$ guarantees to traverse all combinations of $N+1$ adjacent edges, so all prime paths will be traversed as well in a test set satisfying this \textit{N-Switch Coverage} criterion because these prime paths are not longer than $N$.





\textbf{Boundary-interior Coverage subsumes Edge Coverage.}
\textit{Explanation:} Since \textit{Boundary-Interior Coverage} only reduces the number of repetitions of cycles within the traversal of the graph, it covers all edges as well as \textit{All Paths Coverage} do. 


In this overview, we have discussed the relations between coverage criteria listed in Section \ref{sec:coverage_criteria_for_graph} that are necessary to construct the hierarchy graph presented in Figure \ref{fig:criteria_hierarchy}. Some relations are not explicitly mentioned since they are a result of the transitivity of the subsume relation. An overview of these relations is also given in Table \ref{tab:graph_based_criterion_hierarchy_overview_table}, using the following notation:

\begin{description}
\item[S] - Criterion in a row subsumes a criterion in a column.
\item[E] - Criterion in a row is equal to a criterion in a column.
\item[I] - Criterion in a row is incomparable with a criterion in a column.
\item[N] - Criterion in a row was not compared with a criterion in a column.
\item[!S] - Criterion in a row does not subsume a criterion in a column.
\end{description}

\begin{table*}
\caption{Overview of examined test coverage criteria hierarchy relations.}
\centering
\begin{tabular}{|c|c|c|c|c|c|c|c|c|c|c|c|c|c|}
\hline
& \textbf{NC} & \textbf{EC} & \textbf{BC} & \textbf{EPC} & \textbf{PPC} & \textbf{BPC} & \textbf{SRTC} & \textbf{CRTC} & \textbf{WMC} & \textbf{APC} & \textbf{BIC} \\
\hline
\textbf{NC} &  & !S & !S & !S & !S & !S & I & I & !S & !S & !S  \\
\hline
\textbf{EC} & S &  & E & !S & !S & !S & I & I & !S & !S & !S \\
\hline
\textbf{BC} & S & E &  & !S & !S & !S & I & I & !S & !S & !S \\
\hline
\textbf{EPC} & S & S & S &  & I & I & N & N & I or S & !S & N \\
\hline
\textbf{PPC} & S & S & S & I &  & I or S & S & S & I or S & !S & N  \\
\hline
\textbf{BPC} & S & S & S & I & !S &  & N & N & I or S & !S & N \\
\hline
\textbf{SRTC} & I & I & I & N & !S & N &  & !S & N & !S & N  \\
\hline
\textbf{CRTC} & I & I & I & N & !S & N & S &  & N & !S & N  \\
\hline
\textbf{WMC} & S & S & S & !S & !S & !S & N & N &  & !S & N \\
\hline
\textbf{APC} & S & S & S & S & S & S & S & S & S &  & S \\
\hline
\textbf{BIC} & S & S & S & N & N & N & N & N & N & !S &   \\
\hline
\end{tabular}
\label{tab:graph_based_criterion_hierarchy_overview_table}
\end{table*}

In a few places of Table \ref{tab:graph_based_criterion_hierarchy_overview_table}, there is an "I or S" state. These particular relations that were not necessary to construct the hierarchy graph presented in Figure \ref{fig:criteria_hierarchy} and to determine the exact relation, more analysis has to be done, which will be a subject of the future work.

\section{Related work}
\label{sec:related_work}

In the Model-based Testing field, individual definitions of test coverage criteria are present in the literature, usually in the works presenting test case generation algorithms, e.g., \cite{ModelBasedTestCasesGenerationForOnboardSystem,khalil2017FiniteStateMachineTestingCompleteRoundTripVersusTransitionTrees,bures2017prioritized,bures2015pctgen}. However, less work exists regarding the comparison of individual test coverage criteria and in these works usually small sets of coverage criteria are compared, e.g., \cite{Watson1996StructuredTA,5676826,1245299,792624,Mukherjee2016,Lun2019}. In the general path-based testing, overlapping with FSM-based testing, some basic tests were compared by Ammann and Offutt \cite{ammann2016introduction}.


Watson and McCabe analyze Basis Path Coverage and experimentally proved that it subsumes Branch Coverage and Edge Coverage \cite{Watson1996StructuredTA}.

Lun and Chi \cite{5676826} discussed a set of path-based testing coverage criteria for software architecture testing and explained their mutual subsumption formally. Further, Lun, Chi and Xu propose another two path-based coverage criteria for component testing, Node-Sequence Coverage and Edge-sequence Coverage, and prove the subsumption relationships altogether with criteria from their previous work  \cite{5676826,Lun2019}.

Marré and Bertolino have indirectly focused on coverage criteria subsumption when analyzing subsumptions of requirements which are sub-parts of directed graph-based SUT model that need to be covered to satisfy given criterion and are specific for each model \cite{1245299}.

Sinha and Harrold analyzed coverage criteria for data-flow testing using a directed-graph based SUT model and proposed a hierarchy where various criteria as All-du-paths, All-uses, All-p-uses/some-c-uses, All-c-uses/some-p-uses, All-p-uses, All-c-uses, All-defs and general graph-based SUT model criteria as All-path Coverage, Edge Coverage and Node Coverage are compared \cite{792624}.

As we have not found a comprehensive overview of test coverage criteria dedicated directly to FSM testing, we provided such an overview in this paper.

\section{Conclusion}
\label{sec:conclusion}

In this paper, we have summarized the 14 most common test coverage criteria for FSM-based test case generation, including seven synonyms, discussed relations among these criteria, and put them in a hierarchy regarding their subsumption. For each of the discussed relations between individual criteria, an explanation documenting the relationship is given. Part of the discussed criteria is common for general path-based testing based on directed graphs as a system under test model as well as for finite state machines specifically (e.g., Node Coverage, Edge Coverage). This overlap is natural and is a result of the same basic structure, $\mathcal{G}$ used as a SUT model. It is necessary to mention that more alternative coverage criteria exist and are significantly overlapping with the criteria listed in section \ref{sec:coverage_criteria_for_graph}. As an example, we can give an overview by Souza et al., which lists more alternatives \cite{de2017h}; however, due to our opinion and industrial praxis, not so common. For this reason, we decided not to include them in this overview. 

The presented overview is helpful for researchers and various industrial test engineers, as FSMs are one of the most common and widely-used modeling notations in the generation of test cases for software, electronics, or IoT system testing.

\section*{Acknowledgment}

The project is supported by CTU in Prague internal grant SGS20/177/OHK3/3T/13 “Algorithms and solutions for automated generation of test scenarios for software and IoT systems.” The authors acknowledge the support of the OP VVV funded project CZ.02.1.01/0.0/0.0/16\_019 /0000765 “Research Center for Informatics.” Bestoun S. Ahmed has been supported by the Knowledge Foundation of Sweden (KKS) through the Synergi Project AIDA - A Holistic AI-driven Networking and Processing Framework for Industrial IoT (Rek:20200067).

\bibliographystyle{IEEEtran}
\bibliography{REFERENCES}

\begin{thebibliography}{10}
\providecommand{\url}[1]{#1}
\csname url@samestyle\endcsname
\providecommand{\newblock}{\relax}
\providecommand{\bibinfo}[2]{#2}
\providecommand{\BIBentrySTDinterwordspacing}{\spaceskip=0pt\relax}
\providecommand{\BIBentryALTinterwordstretchfactor}{4}
\providecommand{\BIBentryALTinterwordspacing}{\spaceskip=\fontdimen2\font plus
\BIBentryALTinterwordstretchfactor\fontdimen3\font minus
  \fontdimen4\font\relax}
\providecommand{\BIBforeignlanguage}[2]{{%
\expandafter\ifx\csname l@#1\endcsname\relax
\typeout{** WARNING: IEEEtran.bst: No hyphenation pattern has been}%
\typeout{** loaded for the language `#1'. Using the pattern for}%
\typeout{** the default language instead.}%
\else
\language=\csname l@#1\endcsname
\fi
#2}}
\providecommand{\BIBdecl}{\relax}
\BIBdecl

\bibitem{ammann2016introduction}
P.~Ammann and J.~Offutt, \emph{Introduction to software testing}.\hskip 1em
  plus 0.5em minus 0.4em\relax Cambridge University Press, 2016.

\bibitem{yin2012real}
Y.~Yin, B.~Liu, and H.~Ni, ``Real-time embedded software testing method based
  on extended finite state machine,'' \emph{Journal of Systems Engineering and
  Electronics}, vol.~23, no.~2, pp. 276--285, 2012.

\bibitem{bosik1991finite}
B.~S. Bosik and M.~{\"U}. Uyar, ``Finite state machine based formal methods in
  protocol conformance testing: from theory to implementation,'' \emph{Computer
  Networks and ISDN Systems}, vol.~22, no.~1, pp. 7--33, 1991.

\bibitem{tanabe2020model}
K.~Tanabe, Y.~Tanabe, and M.~Hagiya, ``Model-based testing for mqtt
  applications,'' in \emph{Joint Conference on Knowledge-Based Software
  Engineering}.\hskip 1em plus 0.5em minus 0.4em\relax Springer, 2020, pp.
  47--59.

\bibitem{wang2019formal}
J.~Wang, \emph{Formal Methods in Computer Science}.\hskip 1em plus 0.5em minus
  0.4em\relax CRC Press, 2019.

\bibitem{hopcroft2001introduction}
J.~E. Hopcroft, R.~Motwani, and J.~D. Ullman, ``Introduction to automata
  theory, languages, and computation,'' \emph{Acm Sigact News}, vol.~32, no.~1,
  pp. 60--65, 2001.

\bibitem{Watson1996StructuredTA}
A.~Watson, T.~McCabe, and D.~Wallace, ``Structured testing: A testing
  methodology using the cyclomatic complexity metric,'' 1996.

\bibitem{5676826}
L.~Lun and X.~Chi, ``Investigating subsumption relationship on path coverage
  criteria for software architecture testing,'' in \emph{2010 International
  Conference on Computational Intelligence and Software Engineering}, 2010, pp.
  1--4.

\bibitem{1245299}
M.~Marre and A.~Bertolino, ``Using spanning sets for coverage testing,''
  \emph{IEEE Transactions on Software Engineering}, vol.~29, no.~11, pp.
  974--984, 2003.

\bibitem{792624}
S.~Sinha and M.~Harrold, ``Criteria for testing exception-handling constructs
  in java programs,'' in \emph{Proceedings IEEE International Conference on
  Software Maintenance - 1999 (ICSM'99). 'Software Maintenance for Business
  Change' (Cat. No.99CB36360)}, 1999, pp. 265--274.

\bibitem{Mukherjee2016}
\BIBentryALTinterwordspacing
D.~Mukherjee and R.~Mall, \emph{\BIBforeignlanguage{English}{An investigation
  into effective test coverage}}, ser. Advances in Intelligent Systems and
  Computing, 2016, vol. 396, cited By :2. [Online]. Available:
  \url{www.scopus.com}
\BIBentrySTDinterwordspacing

\bibitem{Lun2019}
\BIBentryALTinterwordspacing
L.~Lun, X.~Chi, and H.~Xu, ``\BIBforeignlanguage{English}{Coverage criteria for
  component path-oriented in software architecture},''
  \emph{\BIBforeignlanguage{English}{Engineering Letters}}, vol.~27, no.~1, pp.
  40--52, 2019, cited By :4. [Online]. Available: \url{www.scopus.com}
\BIBentrySTDinterwordspacing

\bibitem{mariano2019comparing}
M.~M. Mariano, {\'E}.~F. de~Souza, A.~T. Endo, and N.~L. Vijaykumar,
  ``Comparing graph-based algorithms to generate test cases from finite state
  machines,'' \emph{Journal of Electronic Testing}, vol.~35, no.~6, pp.
  867--885, 2019.

\bibitem{ModelBasedTestCasesGenerationForOnboardSystem}
J.~Lv, K.~Li, G.~Wei, T.~Tang, C.~Li, and W.~Zhao, ``Model-based test cases
  generation for onboard system,'' in \emph{2013 IEEE Eleventh International
  Symposium on Autonomous Decentralized Systems (ISADS)}, 2013, pp. 1--6.

\bibitem{CoverageCriteriaForStateTransitionTestingAndModelCheckerBasedTestCaseGeneration}
C.~{de Souza Carvalho} and T.~{Tsuchiya}, ``Coverage criteria for state
  transition testing and model checker-based test case generation,'' in
  \emph{2014 Second International Symposium on Computing and Networking}, 2014,
  pp. 596--598.

\bibitem{Heimdahl2004TestSuiteReductionForModelBasedTests}
M.~Heimdahl and D.~George, ``Test-suite reduction for model based tests:
  effects on test quality and implications for testing,'' in \emph{Proceedings.
  19th International Conference on Automated Software Engineering, 2004.},
  2004, pp. 176--185.

\bibitem{AnApproachToProgramTesting}
J.~C. Huang, ``An approach to program testing,'' \emph{ACM Comput. Surv.},
  vol.~7, no.~3, p. 113–128, Sep. 1975.

\bibitem{Li2009AnExperimentalComparisonOfFourUnitTestCriteria}
N.~Li, U.~Praphamontripong, and J.~Offutt, ``An experimental comparison of four
  unit test criteria: Mutation, edge-pair, all-uses and prime path coverage,''
  in \emph{2009 International Conference on Software Testing, Verification, and
  Validation Workshops}, 2009, pp. 220--229.

\bibitem{ANewApproachToGeneratingHighQualityTestCases}
P.~{Liu} and H.~{Miao}, ``A new approach to generating high quality test
  cases,'' in \emph{2010 19th IEEE Asian Test Symposium}, 2010, pp. 71--76.

\bibitem{Fazli2019ATimeAndSpaceEfficientCompositionalMethodForPrimeAndTestPathsGeneration}
E.~Fazli and M.~Afsharchi, ``A time and space-efficient compositional method
  for prime and test paths generation,'' \emph{IEEE Access}, vol.~7, pp.
  134\,399--134\,410, 2019.

\bibitem{bures2019employment}
M.~Bures and B.~S. Ahmed, ``Employment of multiple algorithms for optimal
  path-based test selection strategy,'' \emph{Information and Software
  Technology}, vol. 114, pp. 21--36, 2019.

\bibitem{bures2019prioritized}
M.~Bures, B.~S. Ahmed, and K.~Z. Zamli, ``Prioritized process test: An
  alternative to current process testing strategies,'' \emph{International
  Journal of Software Engineering and Knowledge Engineering}, vol.~29, no.~07,
  pp. 997--1028, 2019.

\bibitem{StateBasedTestsSuitesAutomaticGenerationTool}
H.~Khalil and Y.~Labiche, ``State-based tests suites automatic generation tool
  (stage-1),'' in \emph{2017 IEEE 41st Annual Computer Software and
  Applications Conference (COMPSAC)}, vol.~1, 2017, pp. 357--362.

\bibitem{khalil2017OnFSMBasedTesting}
------, ``On fsm-based testing: An empirical study: Complete round-trip versus
  transition trees,'' in \emph{2017 IEEE 28th International Symposium on
  Software Reliability Engineering (ISSRE)}, 2017, pp. 305--315.

\bibitem{khalil2017FiniteStateMachineTestingCompleteRoundTripVersusTransitionTrees}
H.~Khalil, ``Finite state machine testing complete round-trip versus transition
  trees: On the road of finding the most effective criterion,'' in \emph{2017
  IEEE International Symposium on Software Reliability Engineering Workshops
  (ISSREW)}, 2017, pp. 108--111.

\bibitem{McCabe1976AComplexityMeasure}
T.~McCabe, ``A complexity measure,'' \emph{IEEE Transactions on Software
  Engineering}, vol. SE-2, no.~4, pp. 308--320, 1976.

\bibitem{ZAPATA2013256}
F.~Zapata, A.~Akundi, R.~Pineda, and E.~Smith, ``Basis path analysis for
  testing complex system of systems,'' \emph{Procedia Computer Science},
  vol.~20, pp. 256--261, 2013, complex Adaptive Systems.

\bibitem{Watson1996StructuredTA2}
A.~Watson, ``Structured testing: analysis and extensions,'' 1996.

\bibitem{Wang2019BasisPathCoverageCriteriaForSmartContractApplicationTesting}
X.~Wang, Z.~Xie, J.~He, G.~Zhao, and R.~Nie, ``Basis path coverage criteria for
  smart contract application testing,'' in \emph{2019 International Conference
  on Cyber-Enabled Distributed Computing and Knowledge Discovery (CyberC)},
  2019, pp. 34--41.

\bibitem{TestingSoftwareDesignModeledByFiniteStateMachines}
T.~S. {Chow}, ``Testing software design modeled by finite-state machines,''
  \emph{IEEE Transactions on Software Engineering}, vol. SE-4, no.~3, pp.
  178--187, 1978.

\bibitem{vroon2013tmap}
M.~Vroon, B.~Broekman, T.~Koomen, and L.~van~der Aalst, \emph{TMap next: for
  result-driven testing}.\hskip 1em plus 0.5em minus 0.4em\relax Uitgeverij
  kleine Uil, 2013.

\bibitem{pol2002software}
M.~Pol, R.~Teunissen, and E.~Van~Veenendaal, \emph{Software testing: A guide to
  the TMap approach}.\hskip 1em plus 0.5em minus 0.4em\relax Pearson Education,
  2002.

\bibitem{aalst2008tmap}
L.~v.~d. Aalst, R.~Baarda, E.~Roodenrijs, J.~Vink, and B.~Visser, ``Tmap next,
  business driven test managament,'' 2008.

\bibitem{ConcurrentNSwitchCoverageCriterionForGeneratingTestCasesFromPlaceTransitionNets}
T.~{Takagi}, N.~{Oyaizu}, and Z.~{Furukawa}, ``Concurrent n-switch coverage
  criterion for generating test cases from place/transition nets,'' in
  \emph{2010 IEEE/ACIS 9th International Conference on Computer and Information
  Science}, 2010, pp. 782--787.

\bibitem{Howden1975MethodologyForTheGenerationOfProgramTestData}
W.~Howden, ``Methodology for the generation of program test data,'' \emph{IEEE
  Transactions on Computers}, vol. C-24, no.~5, pp. 554--560, 1975.

\bibitem{GenerationOfIntegrationTestsForSelf-TestingComponents}
L.~Mariani, M.~Pezz{\`e}, and D.~Willmor, ``Generation of integration tests for
  self-testing components,'' in \emph{Applying Formal Methods: Testing,
  Performance, and M/E-Commerce}, M.~N{\'u}{\~{n}}ez, Z.~Maamar, F.~L. Pelayo,
  K.~Pousttchi, and F.~Rubio, Eds.\hskip 1em plus 0.5em minus 0.4em\relax
  Berlin, Heidelberg: Springer Berlin Heidelberg, 2004, pp. 337--350.

\bibitem{bures2017prioritized}
M.~Bures, T.~Cerny, and M.~Klima, ``Prioritized process test: More efficiency
  in testing of business processes and workflows,'' in \emph{International
  Conference on Information Science and Applications}.\hskip 1em plus 0.5em
  minus 0.4em\relax Springer, 2017, pp. 585--593.

\bibitem{bures2015pctgen}
M.~Bures, ``Pctgen: Automated generation of test cases for application
  workflows,'' in \emph{New Contributions in Information Systems and
  Technologies}.\hskip 1em plus 0.5em minus 0.4em\relax Springer, 2015, pp.
  789--794.

\bibitem{de2017h}
{\'E}.~F. De~Souza, V.~A. de~Santiago~J{\'u}nior, and N.~L. Vijaykumar,
  ``H-switch cover: a new test criterion to generate test case from finite
  state machines,'' \emph{Software Quality Journal}, vol.~25, no.~2, pp.
  373--405, 2017.

\end{thebibliography}

\end{document}